\documentclass[showpacs,aps,twocolumn]{revtex4}
\usepackage{adjustbox}
\usepackage{epsfig}
\usepackage{graphicx}
\usepackage{amsmath,amssymb,amsfonts}
\usepackage{array}
\usepackage{url}
\usepackage{hyperref}
\usepackage{multirow}
\usepackage{float}
\usepackage{lineno}
\usepackage{xspace}
\usepackage[usenames,dvipsnames]{color}

\newcommand{\bef}{\begin{figure}}
\newcommand{\eef}{\end{figure}}
\newcommand{\bc}{\begin{center}}
\newcommand{\ec}{\end{center}}

\newcommand{\be}{\begin{equation}}
\newcommand{\ee}{\end{equation}}
\newcommand{\bea}{\begin{eqnarray}}
\newcommand{\eea}{\end{eqnarray}}

\begin{document}

\title{Deciphering QCD dynamics in small collision systems using event shape and final state multiplicity at the Large Hadron Collider}
\author{Suman Deb}
\author{Sushanta Tripathy\footnote{Presently at: Instituto de Ciencias Nucleares, UNAM, Deleg. Coyoac\'{a}n, Ciudad de M\'{e}xico 04510
}}
\author{Golam Sarwar}
\author{Raghunath Sahoo\footnote{Corresponding Author Email: Raghunath.Sahoo@cern.ch}}
\affiliation{Discipline of Physics, Indian Institute of Technology Indore, Simrol, Indore 453552, India}
\author{Jan-e Alam}
\affiliation{Variable Energy Cyclotron Centre, 1/AF, Bidhan Nagar, Kolkata - 700064, India}

\begin{abstract}
The high-multiplicity pp collisions at the Large Hadron Collider energies with various heavy-ion-like signatures have warranted a deeper understanding of the underlying physics and particle production mechanisms. It is a common practice to use experimental data on the  hadronic 
transverse momentum ($p_T$) spectra to extract thermodynamical properties of the system formed in heavy ion and high multiplicity pp collisions.  The non-availability of event topology dependent experimental data for pp collisions at $\sqrt{s}$ = 13 TeV  on the  spectra of non-strange and strange hadrons constrains us to use the PYTHIA8 simulated numbers to extract  temperature-like parameters to study the event shape and multiplicity dependence of specific heat  capacity, conformal symmetry breaking measure (CSBM) and speed of sound. The observables show a clear dependence on event multiplicity and event topology. Thermodynamics of the system is largely governed by the light particles because of their relatively larger abundances. In this regards, a threshold in the particle production, $\rm N_{ch} \simeq$ (10-20) in the final state multiplicity emerges out from the present study, confirming some of the earlier findings in this direction. As for heavier hadrons with relatively small abundances, a similar threshold is observed for $\langle \rm N_{ch} \rangle \simeq$ 40 hinting towards formation of a thermal bath where all the heavier hadrons are in equilibrium.
\end{abstract}
\date{\today}
\maketitle

\section{Introduction}
\label{intro}

To reveal the nature of the Quantum Chromodynamics (QCD) phase transition and to get a glimpse of how matter behaves at extreme conditions of temperature and energy density, experiments like Relativistic Heavy Ion Collider (RHIC) at BNL, USA and Large Hadron
Collider (LHC) at CERN, Geneva, Switzerland have got prime importance. A deconfined state of quarks and gluons, also known as Quark Gluon Plasma (QGP), is believed to be produced for very short lifetime in heavy-ion collisions in these experiments. In the QGP phase, the relevant degrees of freedom are quarks and gluons rather than mesons and baryons, which are confined color neutral states~\cite{Aoki:2006we}. 
In a baffling development, the experiments at LHC discovered QGP-like properties such as strangeness enhancement~\cite{ALICE:2017jyt}, double-ridge structure~\cite{Khachatryan:2016txc} etc. in smaller collision systems like proton-proton (pp) and p-Pb collisions. These developments have important consequences on the results obtained from heavy-ion collisions as pp collisions so far have been used as a benchmark for heavy-ion collisions to understand a possible medium formation. To properly characterise QGP-like properties and to understand origin of such observations in small systems, a comprehensive understanding should be made on the nature of the medium formation by constituents whose interactions are governed by QCD.  Ideally, the QCD thermodynamics can be specified by temperature (T) and 
chemical potential ($\mu$). 
The phase transition to QGP is characterized by change in thermodynamic quantities such as pressure, entropy and energy density at the transition point. Also, it can be characterized using a set of response functions like, specific heat, isothermal compressibility, speed of sound etc. with change in temperature and chemical potential~\cite{Basu:2016ibk}. As an effort to have a better understanding in this direction, we attempt to study the event shape and multiplicity dependence of specific heat capacity, conformal symmetry breaking measure (CSBM) and speed of sound in pp collisions at $\sqrt{s}$ = 13 TeV using PYTHIA8~\cite{pythia8}, one of the widely used event generators in the LHC era.

The specific heat ($C_V$) is estimated via temperature fluctuations, which characterizes the equation of the state of the system. The specific heat is expected to diverge near the critical point for a system undergoing a phase transition. The conformal symmetry breaking measure or trace anomaly plays an important role for QCD dynamics and phase transition. The speed of sound ($c_s$) is used to quantify the softest point of the phase transition along with the location of the critical point~\cite{Mukherjee:2017elm}. To describe the particle production mechanism  and the QCD thermodynamics, the statistical models are more useful due to high multiplicities produced in high-energy collisions. Recently, it has been seen that the experimental particle spectra in high-energy hadronic collisions are successfully explained by Tsallis non-extensive statistics~\cite{Wilk:1999dr,Tsallis:1987eu,Tsallis:1999nq,Thakur:2016boy,Sett:2015lja,Bhattacharyya:2015hya,Zheng:2015gaa,Tang:2008ud,De:2014dna}. Earlier, Tsallis non-extensive statistics has been used as initial distribution in Boltzmann Transport Equation to calculate the elliptic flow and nuclear modification factor in heavy-ion collisions~\cite{Tripathy:2017nmo,Tripathy:2016hlg,Tripathy:2017kwb}. Although there are different variants of Tsallis distribution function, we choose to use a thermodynamically consistent form of Tsallis non-extensive statistical distribution function \cite{Cleymans:2011in}, which nicely describes the $p_{\rm T}$-spectra in LHC pp collisions to calculate the specific heat, CSBM and speed of sound for small collision systems. Recently, the variations of these thermodynamic quantities are studied in
comparison to experimental data \cite{Deb:2019yjo}. However, in view of the production dynamics dependent event topology, it is worth 
making a comprehensive study using one of the topological observables, {\it i.e.} transverse spherocity.


Particle production dynamics in high-energy physics has got two domains, namely the hard pQCD sector and the soft
physics domain, which are not necessarily having a sharp boundary. The hard pQCD sector corresponds to 
high momentum transfer processes, whereas the soft domain is governed by 
low momentum. The soft sector 
event topology is isotropic in nature, while the hard or the jet and mini-jet dominated sector is pencil-like. With high-multiplicity events at the LHC in pp collisions and the observation of heavy-ion-like features, it has become a
necessity to look into event shape and multiplicity dependence of various observables and system thermodynamics.
In order to do that, transverse spherocity ($S_0$) is one of the event-shape observables which has given a new direction for underlying events in pp collisions to have further differential study along with charged particle multiplicity as an
event classifier. Recent studies on transverse spherocity at the LHC suggest that, using event shape one can separate the jetty and isotropic events from the average shaped events \cite{Acharya:2019mzb,Acharya:2018smu,Ortiz:2019osu}. Recently, the chemical and kinetic freeze-out scenario and system thermodynamics are studied using event shape and multiplicity in pp collisions at $\sqrt{s}$ = 13 TeV using PYTHIA8 event generator~\cite{Tripathy:2019blo,Khuntia:2018qox,Rath:2019izg}. These are some of the efforts to understand the
underlying production dynamics and thus paving a way for further experimental investigations in the direction of
possible search for QGP-droplets in small collision systems \cite{Sahoo:2020cle,Sahoo:2019ifs}. For the present analysis, we use the values of temperature and non-extensive parameter, $q$ obtained from Ref.~\cite{Tripathy:2019blo}, where identified particle spectra are
well described by experimentally motivated and thermodynamically consistent Tsallis distribution function.
  
The paper is organized as follows. In section 2, the detailed formalism for specific heat, conformal symmetry breaking measure and speed of sound in non-extensive statistics is introduced. In third section, the details of event generation and analysis methodology are discussed. The next section presents the results and discussions. Finally, in the last section we summarize our findings of this investigation.
\section{Formalism}
\label{formalism}

 For a thermodynamic system at temperature ($T$), volume ($V$), entropy density ($s$), heat capacity ($C_V$) and squared speed of sound ($c_s^2$) can be written as
 \bea
 s = \biggl(\frac{\partial P}{\partial T}\biggr)_{V},
\label{eq1}
\eea
 \bea
 C_V= \biggl(\frac{\partial \epsilon}{\partial T}\biggr)_{V},
\label{eq2}
\eea
\bea
 c_s^2= \biggl(\frac{s}{C_V}\biggr)_{V}.
\label{eq3}
\eea
Here, we take the value of baryonic chemical potential as zero 
because the net baryon number is negligibly small at the central 
rapidity region of the matter formed at LHC energies. $\epsilon$ and $P$ are the energy density and pressure, respectively. At kinetic freeze-out, the momentum distribution of the final state particles is frozen. Thus, these thermodynamical quantities could be estimated from the moments of the momentum distribution at the freeze-out. As mentioned before, it is now established from RHIC to the LHC energies, the particle $p_{\rm T}$ spectra are well explained by Tsallis non-extensive statistics. By fitting Tsallis distribution function to $p_{\rm T}$ spectra of identified particles produced in pp collisions, one can extract thermodynamical parameters- temperature ($T$) and non-extensive parameter ($q$) for the system when final state particles decouple from the system and are detected at the detectors. Then, using these $T$ and $q$, one can calculate all the above quantities obtained using thermodynamically consistent Tsallis distribution function, which is given by~\cite{Cleymans}, 
 \bea
f(E) \equiv  \frac{1}{\exp_{q}(\frac{E}{T})}
 \label{eq4}
\eea

where, 
\begin{equation}
  \exp_{q}(\alpha)=
    \begin{cases}
      [1+(q-1)\alpha]^{\frac{1}{q-1}} & if \ \alpha > 0\\
      [1+(1-q)\alpha]^{\frac{1}{1-q}} & if \ \alpha  \leq 0\\
    \end{cases}       
\end{equation}

where $\alpha$ = $E/T$.
It is to be emphasized that the variable $T$ appearing in this formalism obeys the basic and fundamental thermodynamic relations like (as well as Maxwell's thermodynamic relations)
\begin{equation}
T = \left.\frac{\partial U}{\partial S}\right|_{N,V} ,
\end{equation}
where, $U$ is the internal thermal energy and hence, the parameter $T$ can be called a temperature, even though a system obeys Tsallis and not Boltzmann-Gibbs (BG) statistics.
Using the above distribution function, the mathematical form for the energy density ($\epsilon$), pressure (P), heat capacity ($C_{V}$), conformal symmetry breaking ($\frac{\epsilon-3P}{T^4}$), squared speed of sound ($c_s^{2}$) are:
 
 \bea
  \begin{aligned}
 \epsilon &= \frac{g}{2\pi^2 }\int dp~ p^2 E \times \bigg[1+\frac{(q-1)E}{T}\bigg]^\frac{-q}{q-1},
 \label{eq5}
 \end{aligned}
 \eea
 \bea
 \begin{aligned}
 P &= \frac{g}{2\pi^2 }\int \frac{dp~ p^4}{3E} \times \bigg[1+\frac{(q-1)E}{T}\bigg]^\frac{-q}{q-1},
 \label{eq6}
 \end{aligned}
 \eea
 \bea
  \begin{aligned}
C_V &= \frac{gq}{2\pi^2 T^2}\int dp~ p^2 E^2 \times \bigg[1+\frac{(q-1)E}{T}\bigg]^\frac{1-2q}{q-1},
 \label{eq7}
  \end{aligned}
\eea
\bea
\begin{aligned}
 \frac{\epsilon - 3P}{T^4} &= \frac{g}{2\pi^2 T^4}\int dp~ p^2 E \times\bigg[1-\frac{p^{2}}{E^2}\bigg] \\
                                           &\times\bigg[1+\frac{(q-1)E}{T}\bigg]^\frac{-q}{q-1} ,                                              
\end{aligned}
  \label{eq8}
\eea
\bea
c^2_s =\frac{\frac{gq}{6\pi^2 T^2}\int dp~ p^4 \times \big[1+\frac{(q-1)E}{T}\big]^\frac{1-2q}{q-1}}{C_V}.
  \label{eq9}
\eea
Here, $g$ is the degeneracy factor and $E$ is the energy of the particle given by, $\sqrt{p^2 + m^2}$, where $p$ is the momentum and $m$ is the rest mass of the particle.
\begin{figure}[ht!]
\includegraphics[scale=0.4]{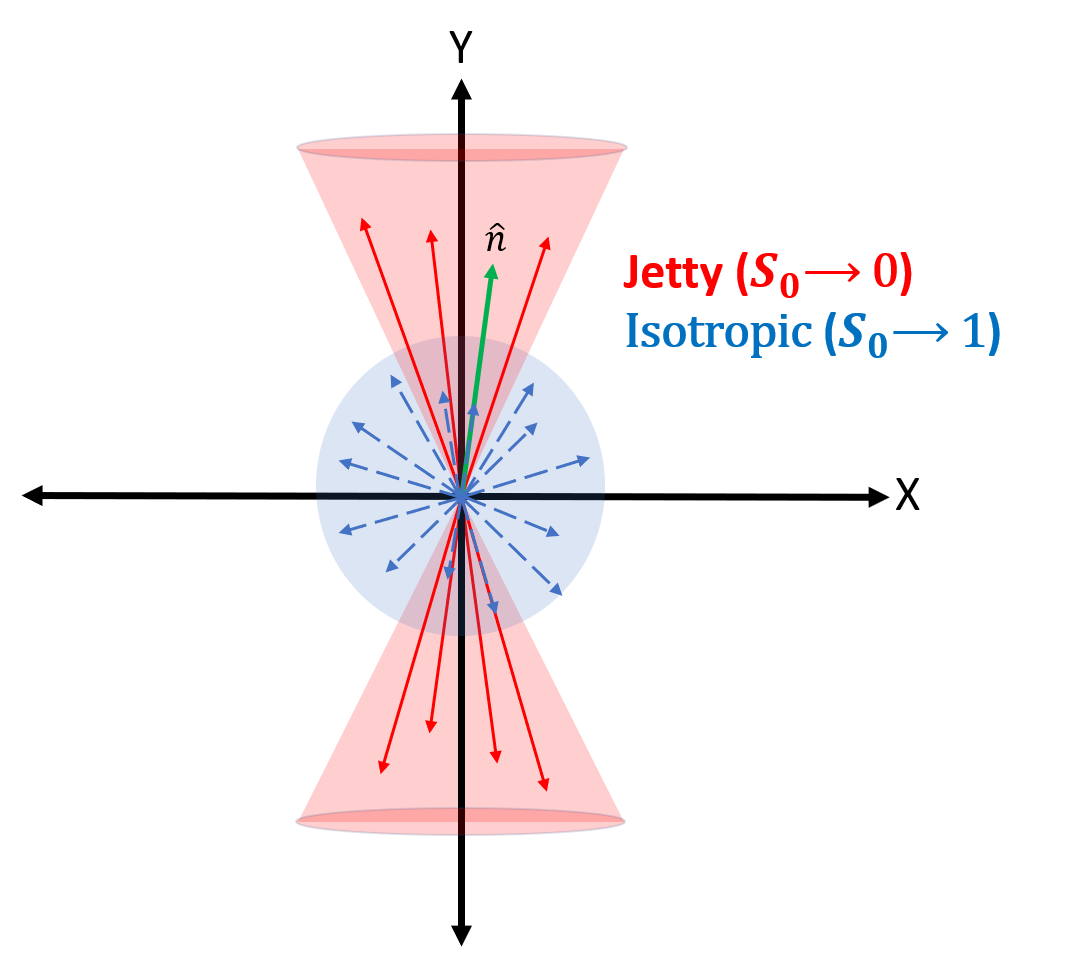}
\caption[]{(Color Online) Schematics of jetty and isotropic events in the transverse plane ($xy$-plane)~\cite{Khuntia:2018qox}.}
\label{sp_cart}
\end{figure}

\begin{figure*}[ht!]
\includegraphics[scale=0.42]{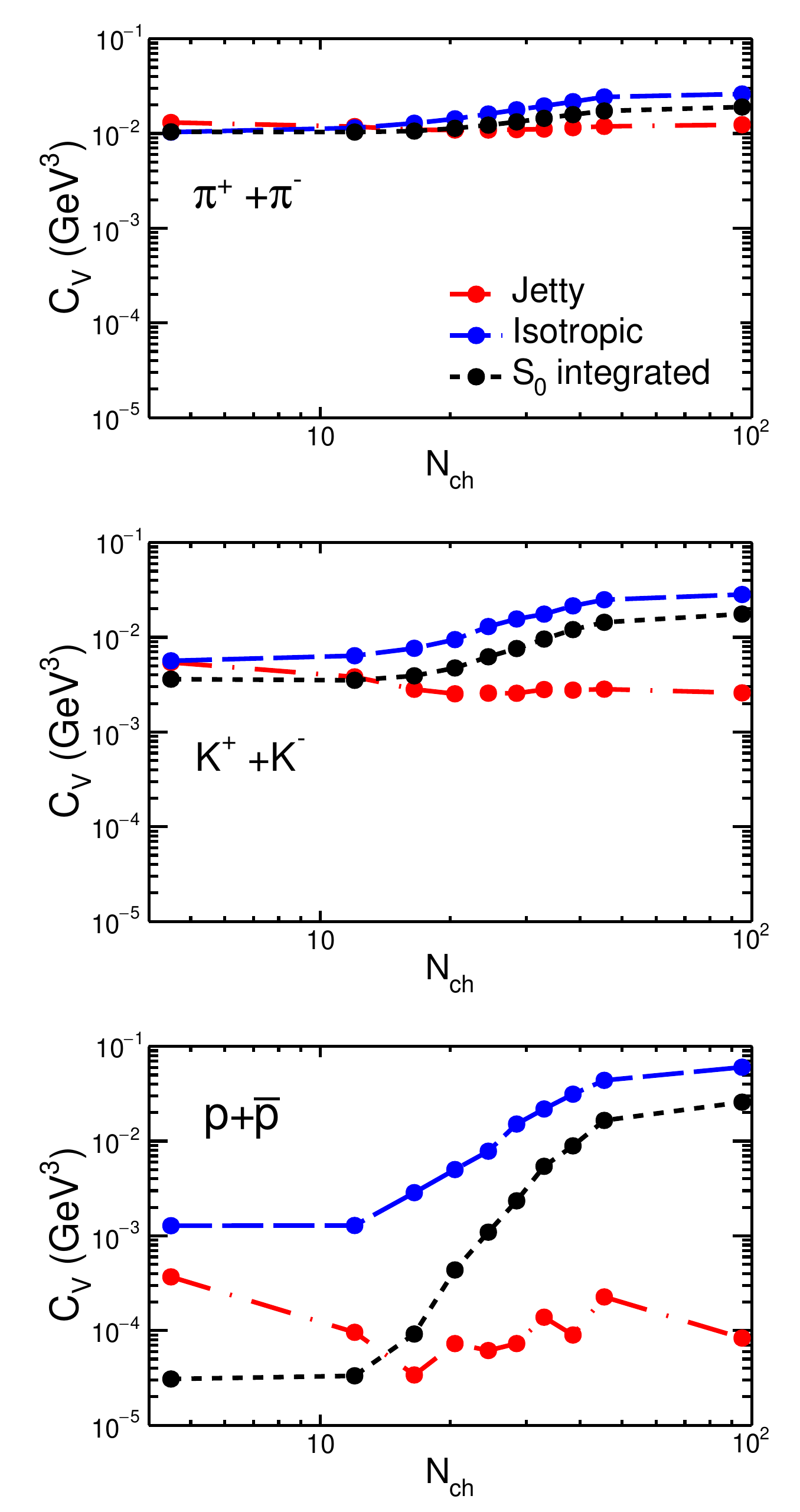}
\includegraphics[scale=0.42]{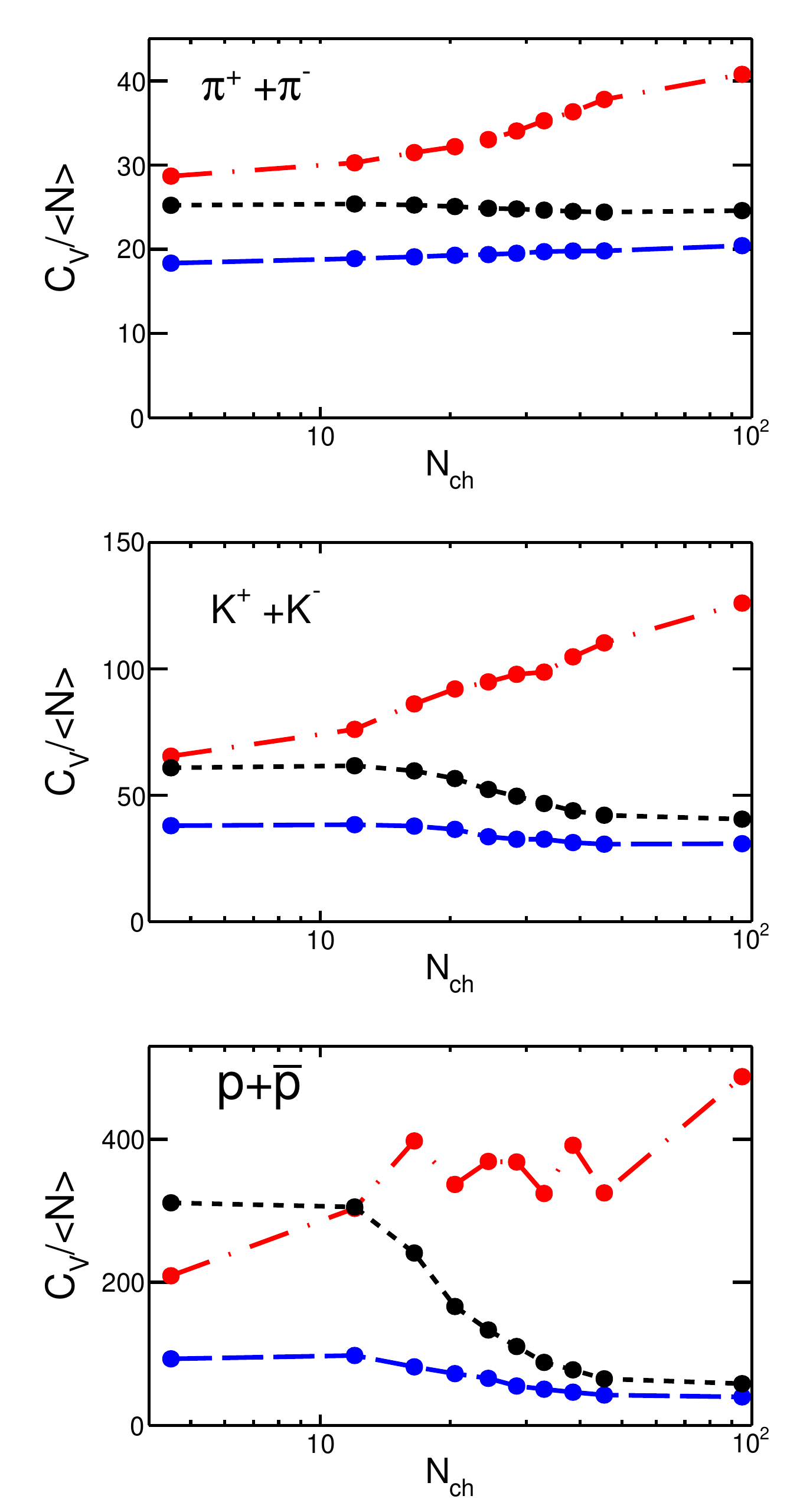}
\includegraphics[scale=0.42]{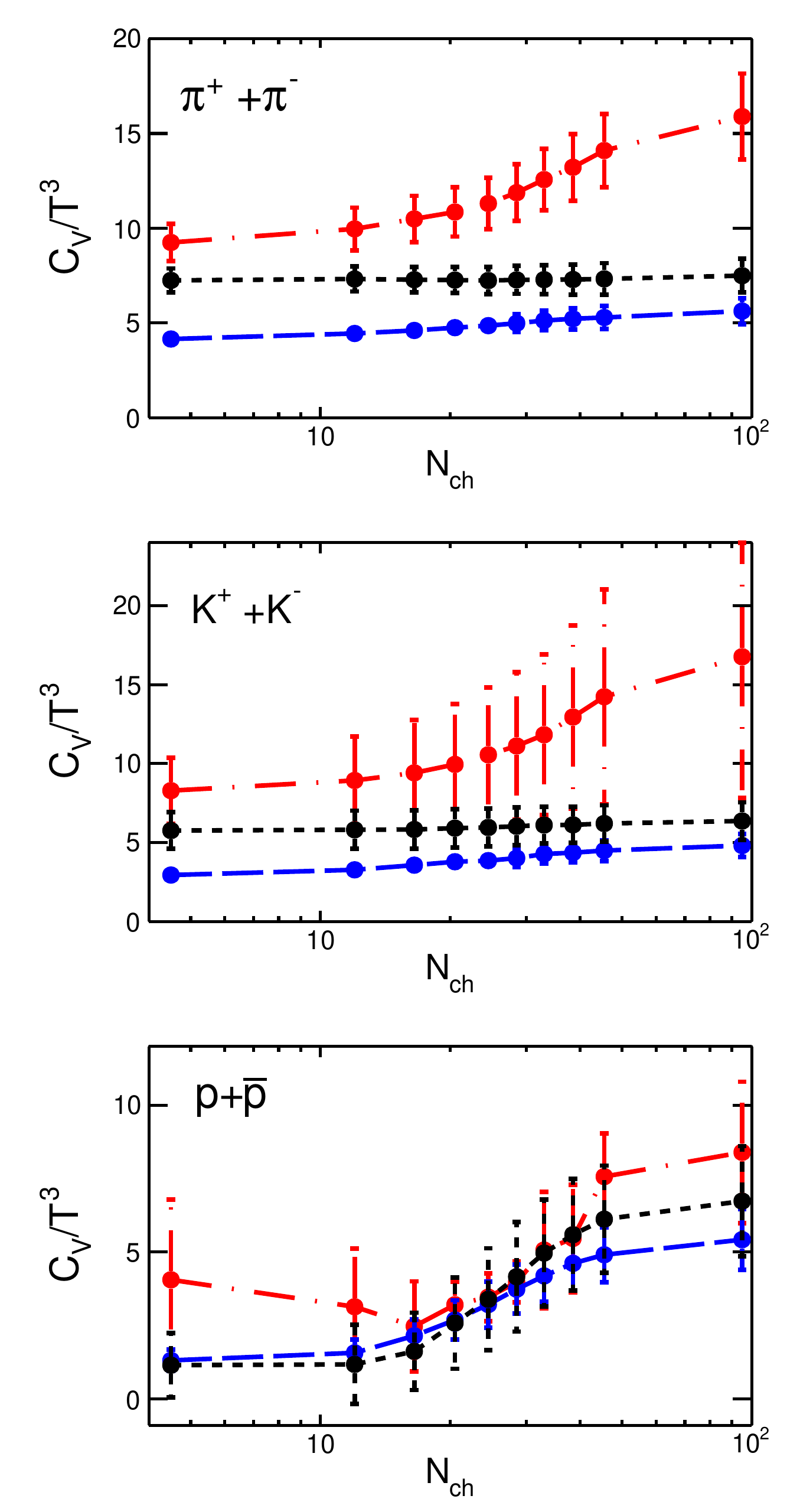}
\caption{(Color Online) Heat Capacity (left), heat capacity scaled by average number of particles (middle) and heat capacity scaled by $T^3$ of the system (right) obtained using Eq.~\ref{eq7} as a function of average charged particle multiplicity for different event shapes for identified light flavor particles.}
 \label{fig2}  
\end{figure*}
\section{Event generation and Analysis methodology}
\label{Event generation}
For this analysis, PYTHIA8 event generator is used to simulate ultra-relativistic pp collisions. It is a blend of many-body physics/theoretical models relevant for hard and soft interactions, initial and final-state parton showers, fragmentation, multipartonic interactions, color reconnection and decay~\cite{Sjostrand:2006za}. We use 8.235 version of PYTHIA~\cite{pythia8}, which includes multipartonic interaction (MPI). MPI is crucial to explain the underlying events, multiplicity distributions and charmonia production \cite{Thakur:2017kpv,Deb:2018qsl,Khatun:2019dml}. Also, this version includes color reconnection which mimics the flow-like effects in pp collisions. 
It is note worthy that PYTHIA8 does not have inbuilt thermalization. However, as discussed in Ref. \cite{Ortiz:2013yxa}, 
the color reconnection (CR) mechanism along with the multipartonic interactions (MPI) in PYTHIA8 produces features those arise from thermalization of a system such as radial flow and mass dependent rise of mean transverse momentum. In the pQCD-based PYTHIA 
model, a single string connecting two partons follows the movement of the partonic endpoints and this movement gives a common boost to the string fragments, which become the final state hadrons. CR along with MPI enables two partons from independent hard scatterings to reconnect and increase the transverse boost. This microscopic treatment of final state particle production is quite similar to a macroscopic picture via hydrodynamical description of high-energy collisions. Thus, it is apparent to conclude that  PYTHIA8 model with MPI and CR, has the ability to mimic the features of thermalization, which is confirmed in the flow-like phenomena in small collision systems  \cite{Ortiz:2013yxa}. This represents a consistent picture because
enhanced MPI leads to thermalization.

We have generated around 250 million pp collision events at $ \sqrt{s}=13~\mathrm{TeV}$ with Monash 2013 Tune (Tune:14)~\cite{Skands:2014pea}. We have implemented the inelastic, non-diffractive component of the total cross-section for all soft QCD processes using the switch SoftQCD:all=on and we use MPI based scheme of color reconnection (ColorReconnection:mode(0)). In our analysis, the minimum bias events are those events where no selection on charged particle multiplicity and spherocity (defined later) is applied. For the generated events, all the hadrons are allowed to decay except the ones used in our study (HadronLevel:Decay = on). Here the event selection criteria is such that only those events were chosen which have at-least 5 tracks (charged particles). The classes based on charged particle multiplicities ($\rm N_{ch}$) have been chosen in the acceptance of V0 detector with pseudorapidity range of V0A ($2.8<\eta<5.1$) and V0C ($-3.7<\eta<-1.7$)~\cite{Abelev:2014ffa} to match with experimental conditions in ALICE at the LHC. The events generated using these cuts are divided in ten multiplicity (V0M) classes, each class containing 10\% of total events, which is tabulated in Table~\ref{tab:V0M}.

\begin{table}[h]
\caption{V0M multiplicity classes and the charged particle multiplicities in each multiplicity class.}
\centering 
\scalebox{0.8}
{
\begin{tabular}{|c|c|c|c|c|c|c|c|c|c|c|} 
\hline 
V0M class & I & II & III & IV & V & VI & VII &VIII & IX & X \\
\hline 
$\rm N_{ch}$ &50-140 & 42-49 & 36-41 & 31-35 & 27-30 & 23-26 & 19-22 &  15-18 & 10-14 & 0-9\\
\hline
\end{tabular}
}
\label{tab:V0M}
\end{table}

Transverse spherocity is defined for an unit vector $\hat{n} (n_{T},0)$ which minimizes the following quantity
\cite{Banfi:2010xy,Cuautle:2014yda}:
\begin{eqnarray}
S_{0} = \frac{\pi^{2}}{4} \bigg(\frac{\Sigma_{i}~|\vec p_{T_{i}}\times\hat{n}|}{\Sigma_{i}~p_{T_{i}}}\bigg)^{2}.
\end{eqnarray}

The events whether they are isotropic or jetty in transverse plane are coupled to the extreme limits of spherocity, which varies from 0 to 1. In the spherocity distribution, the events limiting towards unity are isotropic events while towards zero are jetty ones. The isotropic events are the consequence of soft processes while the jetty events are of hard pQCD processes. Schematically, jetty and isotropic events are shown in Fig.~\ref{sp_cart}~\cite{Khuntia:2018qox}. The spherocity distribution is selected in the pseudorapidity range of $|\eta|<0.8$ to match the experimental conditions of ALICE at the LHC and all events have minimum constraint of 5 charged particles with $p_{\rm{T}}$$>$ 0.15 GeV/$c$  \cite{Acharya:2019mzb}. For minimum bias events (0-100\% V0M class), we consider the jetty events are those having $0\leq S_{0}<0.29$ with lowest 20 percent of total events and the isotropic events are those having $0.64<S_{0}\leq1$ with highest 20 percent of the total events. As shown in our previous work~\cite{Khuntia:2018qox}, spherocity distribution also depends on event multiplicity. Thus, we have considered different spherocity ranges for jetty and isotropic events in different multiplicity classes, which are shown in Table~\ref{table:spherocity range}. With the detailed formalism and analysis methodology, we now move to discuss the results in the next section.

 \begin{table*}[htbp]
\caption{Spherocity ranges for jetty and isotropic events for different multiplicity classes.}
\label{table:spherocity range}
\newcommand{\tabincell}
\centering
\begin{tabular}{|c|c|c|}
\toprule 
\multicolumn{1}{|c|}{\bf V0M Classes}&\multicolumn{2}{c|}{\bf $S_{0}$ range} \\
\cline{2-3}
\multicolumn{1}{|c|}{} &{\bf Jetty events} & {\bf Isotropic events} \\
\hline

\multirow{1}{*}{$0-9$}

&$0-0.20$ &$0.56-1$ \\
\cline{2-3} 
\cline{1-3} 

\multirow{1}{*}{$10-14$}

&0$-$ 0.22 &0.58$-$ 1 \\
\cline{2-3} 
\cline{1-3} 

\multirow{1}{*}{$15-18$}

&0$-$ 0.24&0.60$-$ 1\\
\cline{2-3} 
\cline{1-3} 

\multirow{1}{*}{$19-22$}

&0$-$ 0.26 &0.62$-$ 1 \\

\cline{2-3} 
\cline{1-3} 

\multirow{1}{*}{$23-26$}

&0$-$ 0.28&0.64$-$ 1 \\
\cline{2-3} 
\cline{1-3} 

\multirow{1}{*}{$27-30$}

&0$-$ 0.30 &0.66$-$ 1 \\
\cline{2-3} 
\cline{1-3} 

\multirow{1}{*}{$31-35$}

&0$-$ 0.32 &0.66$-$ 1 \\
\cline{2-3} 
\cline{1-3} 

\multirow{1}{*}{$36-41$} 

&0$-$ 0.34 &0.68$-$ 1 \\
\cline{2-3} 
\cline{1-3} 

\multirow{1}{*}{$42-49$}

&0$-$ 0.38 &0.70$-$ 1 \\
\cline{2-3} 
\cline{1-3} 

\multirow{1}{*}{$50-140$}

&0$-$ 0.42 &0.74$-$ 1 \\
\cline{2-3} 
\cline{1-3}

 \end{tabular}


 \end{table*}

 \begin{table*}[htbp]
\caption{The extracted Tsallis parameters ($T$, $q$) for identified particles in different multiplicity and spherocity classes.}
\label{table:parameters}
\newcommand{\tabincell}
\centering
\begin{adjustbox}{max width=\textwidth}
\begin{tabular}{|c|c|c|c|c|c|c|c|c|c|c|c|c|}\toprule 
\multicolumn{3}{|l|}{${\bf Particles}$}&\multicolumn{10}{c|}{\bf V0M classes} \\
\cline{4-13}
\multicolumn{3}{|c|}{} &{\bf I} & {\bf II} & {\bf III} & {\bf IV}& {\bf V} &{\bf VI} & {\bf VII} & {\bf VIII} & {\bf IX}&{\bf X}\\
\hline


\multirow{6}{*}{$\bf{\pi^+ +\pi^-}$} 

& \multirow{3}{*}{T (GeV)} 


&$\rm S_{0}$-int&0.136 $\pm$ 0.005&0.133 $\pm$ 0.005 &0.129 $\pm$ 0.005&0.126 $\pm$ 0.004&0.122 $\pm$ 0.004&0.119 $\pm$ 0.004 &0.116 $\pm$ 0.004&0.113 $\pm$ 0.003 &0.112 $\pm$ 0.003&0.113 $\pm$ 0.003 \\

\cline{3-13} 


&&Jetty&0.092 $\pm$ 0.004&0.094 $\pm$ 0.004 &0.096 $\pm$ 0.004 &0.096 $\pm$ 0.004&0.097 $\pm$ 0.004&0.099 $\pm$ 0.001&0.100 $\pm$ 0.004&0.101 $\pm$ 0.004 &0.106 $\pm$ 0.004&0.112 $\pm$ 0.004\\

\cline{3-13} 


&&Iso&0.167 $\pm$ 0.007&0.166 $\pm$ 0.006 &0.161 $\pm$ 0.006&0.156 $\pm$ 0.005&0.153 $\pm$ 0.005&0.149 $\pm$ 0.004 &0.144 $\pm$ 0.004&0.141 $\pm$ 0.004 &0.137 $\pm$ 0.003&0.136 $\pm$ 0.003\\

\cline{2-13} 

& \multirow{3}{*}{q} 

&$\rm S_{0}$-int&1.153 $\pm$ 0.001&1.151 $\pm$ 0.003 &1.150 $\pm$ 0.003&1.150 $\pm$ 0.003 &1.150 $\pm$ 0.003 &1.150 $\pm$ 0.003 &1.150 $\pm$ 0.003&1.151 $\pm$ 0.003 &1.151 $\pm$ 0.003& 1.150 $\pm$ 0.003 \\

\cline{3-13}

&&Jetty &1.220 $\pm$ 0.004&1.121 $\pm$ 0.004 &1.1206 $\pm$ 0.004&1.202 $\pm$ 0.004 &1.197 $\pm$ 0.004 &1.193 $\pm$ 0.003 &1.189 $\pm$ 0.003&1.186 $\pm$ 0.003 &1.181 $\pm$ 0.003&1.174 $\pm$ 0.003\\

\cline{3-13} 


&&Iso &1.121 $\pm$ 0.004&1.113 $\pm$ 0.003 &1.112 $\pm$ 0.003&1.110 $\pm$ 0.003 &1.107 $\pm$ 0.003 &1.104 $\pm$ 0.003 & 1.101 $\pm$ 0.003 &1.097 $\pm$ 0.003 &1.093 $\pm$ 0.003 &1.084 $\pm$ 0.003\\

\cline{1-13}


\multirow{6}{*}{$\bf{K^{+}+ K^{-}}$} 

& \multirow{3}{*}{T (GeV)} 


&$\rm S_{0}$-int&0.140 $\pm$ 0.009&0.132 $\pm$ 0.008 &0.125 $\pm$ 0.008&0.116 $\pm$ 0.007&0.108 $\pm$ 0.007 &0.102 $\pm$ 0.007 &0.093 $\pm$ 0.006&0.088 $\pm$ 0.006 &0.085 $\pm$ 0.006&0.086 $\pm$ 0.006 \\

\cline{3-13}  


&&Jetty&0.054 $\pm$ 0.010&0.058 $\pm$ 0.009 &0.060 $\pm$ 0.009 &0.0062 $\pm$ 0.009&0.061 $\pm$ 0.009&0.062 $\pm$ 0.008 & 0.063 $\pm$ 0.008&0.067 $\pm$ 0.008 & 0.075 $\pm$ 0.008&0.087 $\pm$ 0.007 \\

\cline{3-13} 


&&Iso&0.181 $\pm$ 0.009&0.177 $\pm$ 0.009 &0.170 $\pm$ 0.008&0.160 $\pm$ 0.007&0.157 $\pm$ 0.007 &0.150 $\pm$ 0.007 & 0.136 $\pm$ 0.006&0.129 $\pm$ 0.006 &0.125 $\pm$ 0.006&0.124 $\pm$ 0.005\\

\cline{2-13} 

& \multirow{3}{*}{q} 

&$\rm S_{0}$-int&1.155 $\pm$ 0.005&1.155 $\pm$ 0.005 &1.156 $\pm$ 0.004 &1.159 $\pm$ 0.004 &1.161 $\pm$ 0.004&1.162 $\pm$ 0.004 & 1.166 $\pm$ 0.004&1.167 $\pm$ 0.004 &1.169 $\pm$ 0.004 & 1.167 $\pm$ 0.004 \\

\cline{3-13}

&&Jetty &1.244 $\pm$ 0.007&1.234 $\pm$ 0.007 &1.229 $\pm$ 0.009&1.223 $\pm$ 0.006 & 1.220 $\pm$ 0.006 &1.217 $\pm$ 0.006 &1.213 $\pm$ 0.006&1.208 $\pm$ 0.005 & 1.201 $\pm$ 0.005&1.192 $\pm$ 0.005  \\

\cline{3-13} 

&&Iso &1.121 $\pm$ 0.004&1.114 $\pm$ 0.004 &1.113 $\pm$ 0.004&1.113 $\pm$ 0.004 &1.108 $\pm$ 0.004&1.106 $\pm$ 0.004 & 1.110 $\pm$ 0.004 &1.107 $\pm$ 0.004 &1.102 $\pm$ 0.004 &1.092 $\pm$ 0.004 \\

\cline{1-13}


\multirow{6}{*}{$\bf{p + \overline{p}}$} 

& \multirow{3}{*}{T (GeV)} 


&$\rm S_{0}$-int&0.156 $\pm$ 0.014&0.139 $\pm$ 0.014 &0.117 $\pm$ 0.013&0.103 $\pm$ 0.013&0.083 $\pm$ 0.012 &0.069 $\pm$ 0.012 &0.055 $\pm$ 0.011&0.038 $\pm$ 0.010 &0.031 $\pm$ 0.012 &0.030 $\pm$ 0.010\\ 

\cline{3-13} 


&&Jetty&0.021 $\pm$ 0.002&0.031 $\pm$ 0.002 &0.025 $\pm$ 0.003 &0.030 $\pm$ 0.004&0.026 $\pm$ 0.002 &0.026 $\pm$ 0.002 &0.028 $\pm$ 0.002&0.024 $\pm$ 0.005 &0.031 $\pm$ 0.007 &0.045 $\pm$ 0.010\\

\cline{3-13} 


&&Iso&0.223 $\pm$ 0.014&0.207 $\pm$ 0.013 &0.189 $\pm$ 0.013&0.173 $\pm$ 0.012&0.159 $\pm$ 0.012&0.135 $\pm$ 0.011 &0.123 $\pm$ 0.010&0.110 $\pm$ 0.009 & 0.094 $\pm$ 0.009&0.099 $\pm$ 0.009        \\

\cline{2-13} 

& \multirow{3}{*}{q} 


&$\rm S_{0}$-int&1.132 $\pm$ 0.006&1.134 $\pm$ 0.006 &1.141 $\pm$ 0.006&1.144 $\pm$ 0.006 &1.151 $\pm$ 0.006 &1.155 $\pm$ 0.006 & 1.159 $\pm$ 0.005&1.166 $\pm$ 0.005 &1.170 $\pm$ 0.006& 1.170 $\pm$ 0.005 \\

\cline{3-13}


&&Jetty &1.235 $\pm$ 0.002&1.221 $\pm$ 0.003 &1.218 $\pm$ 0.003&1.210 $\pm$ 0.003 &1.208 $\pm$ 0.002 &1.205 $\pm$ 0.002 &1.200 $\pm$ 0.002&1.199 $\pm$ 0.002 &1.194 $\pm$ 0.004&1.186 $\pm$ 0.005\\

\cline{3-13} 

&&Iso &1.089 $\pm$ 0.006&1.087 $\pm$ 0.005 &1.090 $\pm$ 0.005&1.090 $\pm$ 0.005 &1.090 $\pm$ 0.005 &1.097 $\pm$ 0.005 &1.096 $\pm$ 0.005 &1.096 $\pm$ 0.005 &1.098 $\pm$ 0.005&1.084 $\pm$ 0.005\\

\cline{1-13}


\multirow{6}{*}{$\bf{K^{*0} + \overline {K^{*0} }}$}

& \multirow{3}{*}{T (GeV)} 


&$\rm S_{0}$-int&0.163 $\pm$ 0.015&0.143 $\pm$ 0.015 &0.125 $\pm$ 0.014&0.106 $\pm$ 0.013&0.092 $\pm$ 0.013&0.075 $\pm$ 0.012 & 0.058 $\pm$ 0.012&0.040 $\pm$ 0.012 &0.030 $\pm$ 0.004&0.025 $\pm$ 0.002\\ 

\cline{3-13} 


&&Jetty&0.087 $\pm$ 0.017&0.081 $\pm$ 0.015 &0.070 $\pm$ 0.016&0.057 $\pm$ 0.015&0.050 $\pm$ 0.015&0.041 $\pm$ 0.011 & 0.031 $\pm$ 0.011&0.023 $\pm$ 0.003 &0.019 $\pm$ 0.002&0.019 $\pm$ 0.002\\

\cline{3-13} 


&&Iso&0.183 $\pm$ 0.016&0.165 $\pm$ 0.015 &0.147 $\pm$ 0.014&0.125 $\pm$ 0.014&0.114 $\pm$ 0.014&0.095 $\pm$ 0.013 &0.075 $\pm$ 0.013&0.062 $\pm$ 0.012 &0.048 $\pm$ 0.012 &0.027 $\pm$ 0.010\\

\cline{2-13} 

& \multirow{3}{*}{q} 


&$\rm S_{0}$-int&1.144 $\pm$ 0.007&1.148 $\pm$ 0.007 &1.153 $\pm$ 0.007&1.160 $\pm$ 0.006 &1.163 $\pm$ 0.006 &1.168 $\pm$ 0.006 &1.175 $\pm$ 0.006&1.182 $\pm$ 0.006 &1.188 $\pm$ 0.002& 1.192 $\pm$ 0.002\\

\cline{3-13}

&&Jetty &1.189 $\pm$ 0.009&1.185 $\pm$ 0.008 &1.187 $\pm$ 0.008&1.191 $\pm$ 0.008 &1.192 $\pm$ 0.008&1.194 $\pm$ 0.006 &1.198 $\pm$ 0.006&1.202 $\pm$ 0.003 &1.205 $\pm$ 0.002&1.209 $\pm$ 0.00\\

\cline{3-13} 

&&Iso &1.134 $\pm$ 0.007&1.137 $\pm$ 0.007 &1.140 $\pm$ 0.007&1.148 $\pm$ 0.007 &1.148 $\pm$ 0.007&1.154 $\pm$ 0.007 &1.161 $\pm$ 0.007&1.163 $\pm$ 0.007 &1.166 $\pm$ 0.007&1.175 $\pm$ 0.006 \\

\cline{1-13}


\multirow{6}{*}{$\bf{\Lambda^{0} + \overline {\Lambda^{0} }}$}

& \multirow{3}{*}{T (GeV)} 


&$\rm S_{0}$-int&0.167 $\pm$ 0.020&0.143 $\pm$ 0.019 &0.115 $\pm$ 0.019&0.091 $\pm$ 0.019&0.072 $\pm$ 0.017 &0.049 $\pm$ 0.016 &0.029 $\pm$ 0.004&0.021 $\pm$ 0.002 &0.017 $\pm$ 0.001&0.016 $\pm$ 0.001\\

\cline{3-13} 


&&Jetty&0.053 $\pm$ 0.026&0.040 $\pm$ 0.002 &0.026 $\pm$ 0.005&0.027 $\pm$ 0.003&0.020 $\pm$ 0.002&0.017 $\pm$ 0.001 & 0.017 $\pm$ 0.001&0.015 $\pm$ 0.001 &0.014 $\pm$ 0.001&0.016 $\pm$ 0.001\\

\cline{3-13} 


&&Iso&0.196 $\pm$ 0.021&0.167 $\pm$ 0.020 &0.152 $\pm$ 0.020&0.107 $\pm$ 0.021& 0.097 $\pm$ 0.020 &0.059 $\pm$ 0.019 &0.047 $\pm$ 0.007&0.041$\pm$ 0.003 &0.036 $\pm$ 0.002&0.030 $\pm$ 0.002\\

\cline{2-13} 

& \multirow{3}{*}{q} 



&$\rm S_{0}$-int&1.136$\pm$0.008&1.139 $\pm$0.008 &1.147 $\pm$0.008&1.154 $\pm$ 0.008 &1.159 $\pm$ 0.008&1.167 $\pm$0.007 & 1.174 $\pm$ 0.002&1.176 $\pm$ 0.002 & 1.179 $\pm$0.002&1.180 $\pm$0.002\\

\cline{3-13}


&&Jetty &1.195 $\pm$ 0.013&1.194  $\pm$ 0.003  &1.197  $\pm$0.003&1.192  $\pm$ 0.003 &1.192  $\pm$ 0.002 &1.191  $\pm$ 0.002 &1.189  $\pm$ 0.002&1.190  $\pm$ 0.002  &1.192  $\pm$ 0.002&1.193  $\pm$ 0.002\\

\cline{3-13} 

&&Iso &1.122 $\pm$ 0.008&1.128 $\pm$ 0.008 &1.129 $\pm$ 0.008&1.145 $\pm$ 0.009 &1.144 $\pm$ 0.009&1.160 $\pm$ 0.009 &1.159 $\pm$ 0.004&1.157 $\pm$ 0.002 &1.157 $\pm$ 0.002&1.155 $\pm$ 0.002\\

\cline{1-13} 

 \end{tabular}

\end{adjustbox}

 \end{table*}

\section{Results and Discussion} 
In a hydrodynamically expanding scenario, the produced fireball in ultra-relativistic hadronic and nuclear collisions, expands and cools down, 
resulting in a temperature profile as a function of space-time. The spacetime evolution of hadronic and heavy-ion collisions at the LHC energies could be thought of following such an expansion governed by relativistic hydrodynamics. Different identified particles decouple from different 
evolution stages of the fireball because of the different interaction cross sections of the hadrons of different masses. Therefore, the freeze-out temperature in hadronic and heavy-ion collisions should be species dependent.
This means hadrons with smaller cross sections will escape the system earlier than hadrons with larger interaction cross sections. Hence, each hadron species will measure different freeze-out temperature of the system analogous to the cosmological scenario, where different particles go out of equilibrium at different times during the evolution of the universe. In this work, we have considered such a scenario and have evaluated various quantities which are analogous to the thermodynamic quantities such as the heat capacity, scaled heat capacity, CSBM and $c_s^{2}$ at different decoupling points of final state particles from the produced fireball. However, it is to be noted that this work uses simulated data from PYTHIA8 which lack thermalisation in true sense but mimics its features as explained in section~\ref{Event generation}. 

In such a scenario, let's now proceed to calculate the heat capacity (defined by Eq. \ref{eq7}), heat capacity scaled by number density of hadrons ($<N>$ in $\rm GeV^{3}$ obtained by integrating Eq.~\ref{eq4} over three momentum) and scaled with $T^3$ as a function of charged particle multiplicity and transverse spherocity for pp collisions at $\sqrt{s}$=13 TeV generated using PYTHIA8. The temperature parameter, $T$ and non-extensive parameter, $q$ for different event multiplicity and spherocity classes are extracted by fitting Tsallis distribution function to $p_{\rm T}$ spectra of identified particles, which are tabulated in Table \ref{table:parameters}. It is to be noted that the thermalization represents soft physics, therefore, in the present context, the low $p_{\rm T}$ sector of high multiplicity events (to ensure multiple interactions) will have a greater possibility to achieve thermalization. The parameter extracted from the inverse slope of the $p_{\rm T}$ distributions provided by Tsallis distribution may be realistically treated as the temperature for large multiplicity classes. Thus for example from Table \ref{table:parameters}, the isotropic temperature of $\Lambda^{0}$ particles, T $\approx$ 0.196 GeV obtained for the range $\rm N_{ch}$ = (50-140) with mean value, $\rm \langle N_{ch} \rangle$ = 95 may be sensibly considered as the temperature of the system. However, the value of T $\approx$ 0.03 GeV retrieved from the inverse slope of the $p_{\rm T}$ distribution for the range $\rm N_{ch}$ = (0-9) (with mean value,$\rm \langle N_{ch} \rangle$ = 4.5) can not be treated as a realistic value of the temperature of the system. Because in the latter case ($\rm \langle N_{ch} \rangle$ = 4.5), a sufficient number of interactions may not take place to achieve thermalization as opposed to the former case ($\rm \langle N_{ch} \rangle$ = 95). However, for a systematic study as a function of event multiplicity, we have taken all the multiplicity classes.

The $p_{\rm T}$-spectra and the reduced-$\chi^2$ for different event types and multiplicity classes are shown explicitly in Ref.\cite{Tripathy:2019blo}. Using the same formalism, we estimate the conformal symmetry breaking measure and squared speed of sound, defined by Eq.~\ref{eq8} and \ref{eq9} respectively, for different particles as a function of event multiplicity and spherocity. As the inputs to these equations, the values of $T$ and $q$ are extracted after fitting the $p_\mathrm{T}$-spectra from simulated events using Tsallis distribution given by Eq.~\ref{eq4}. It is evident from Ref.~\cite{Tripathy:2019blo} that different particles have different $T$ and $q$ values. Thus, we consider differential freeze-out scenario. Higher mass particles decouple from the system early in time indicating a higher Tsallis temperature parameter. 
These particles are expected to carry more initial non-equilibrium effects. The $q$-value for BG distribution of an equilibrated system is
unity and the observation of $q > 1$ in high-energy hadronic collisions is an indication of the created system being away from equilibrium. In the present study, we take light flavor identified particles like pions ($\pi^{+}+\pi^{-}$), kaons ($\mathrm{K}^{+} + \mathrm{K}^{-}$), protons ($p + \overline{p}$), which have higher abundances in the system and heavier strange/multi-strange particles like K$^{*0}$ ($\mathrm{K}^{*0} + \overline{\mathrm{K}^{*0}}$),  and $\Lambda^{0} ({\Lambda^{0} + \overline {\Lambda^{0} }})$, which have relatively smaller production rates . 

\begin{figure*}[ht!]
\includegraphics[scale=0.42]{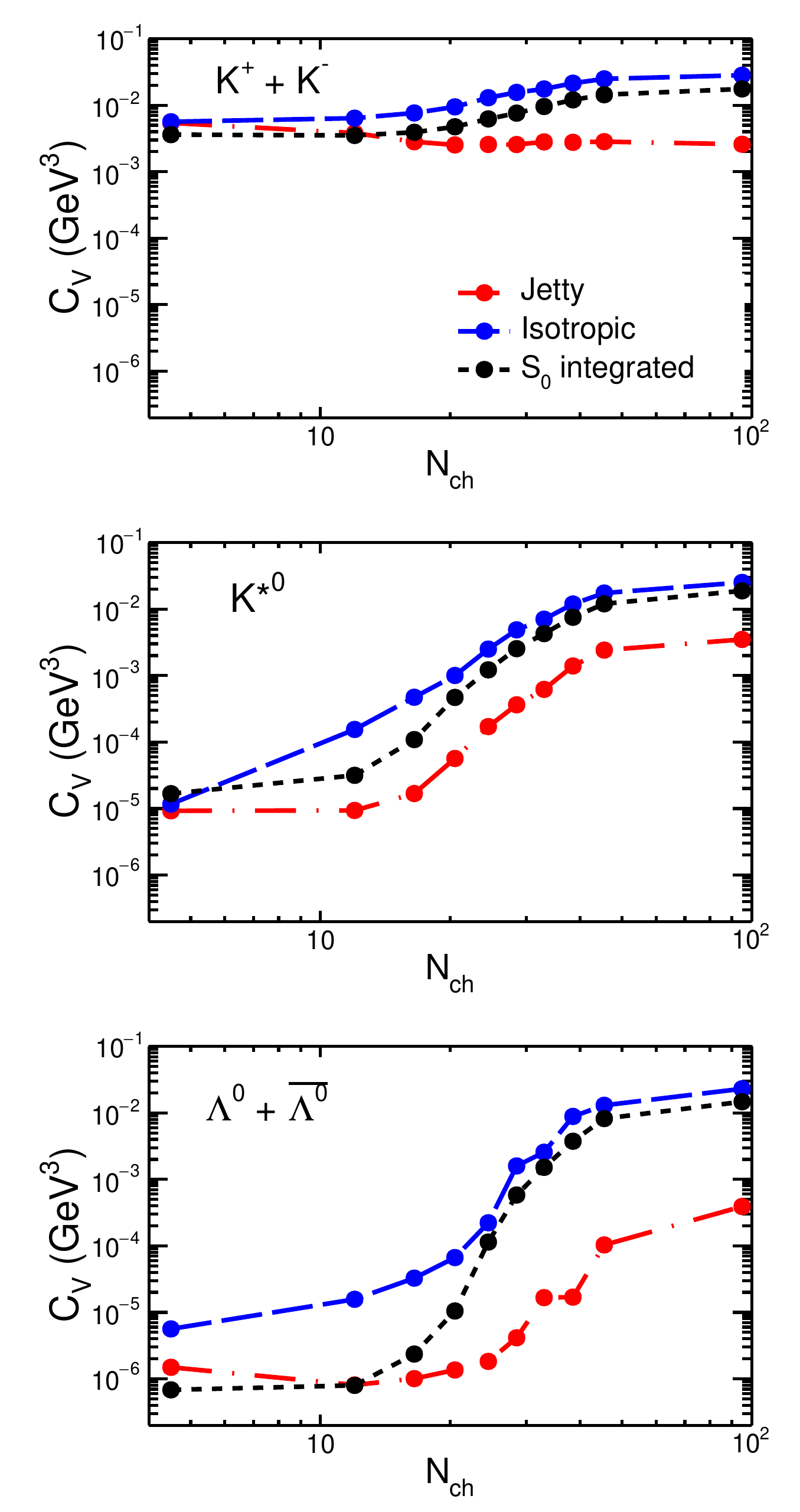}
\includegraphics[scale=0.42]{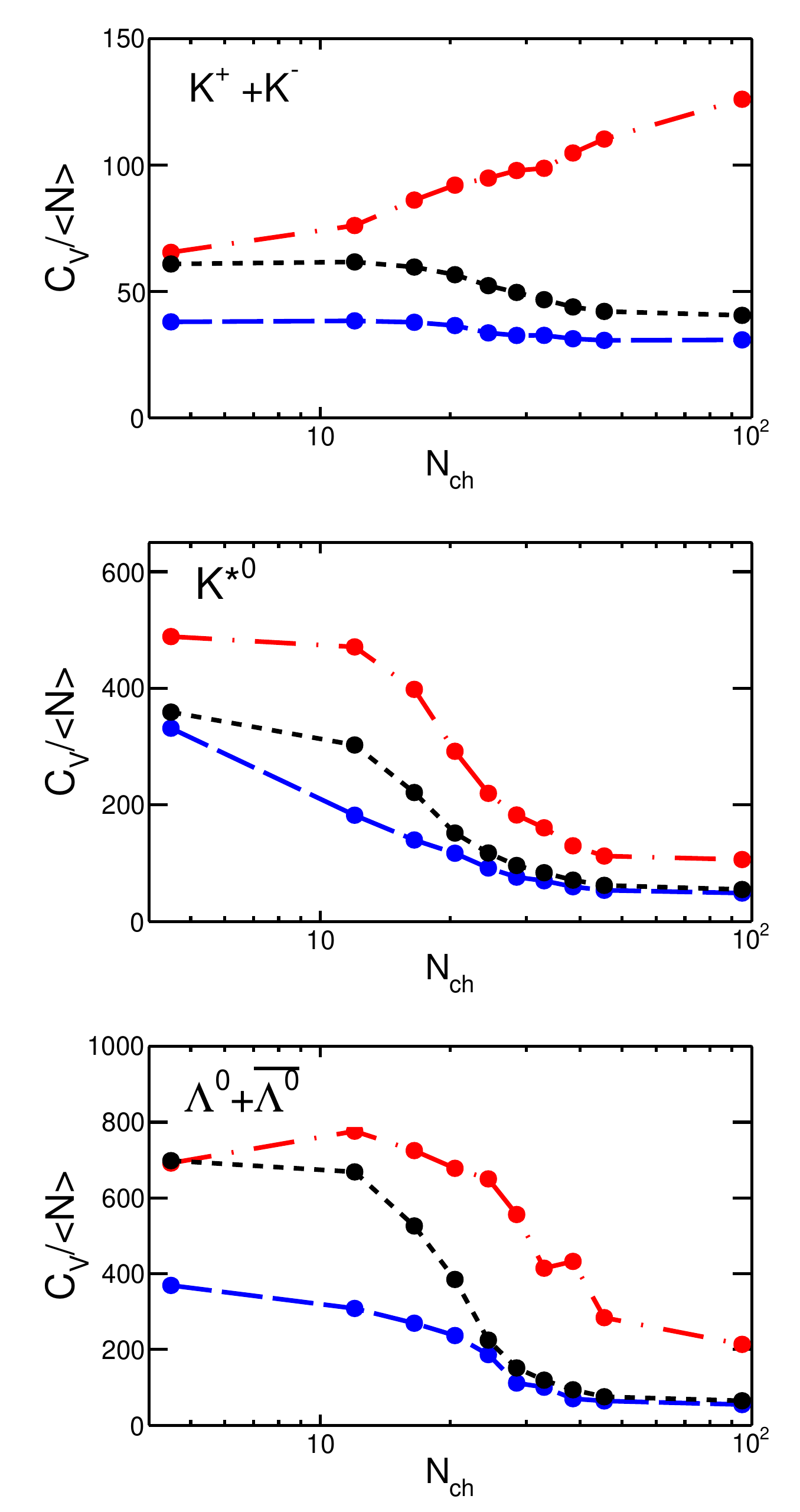}
\includegraphics[scale=0.42]{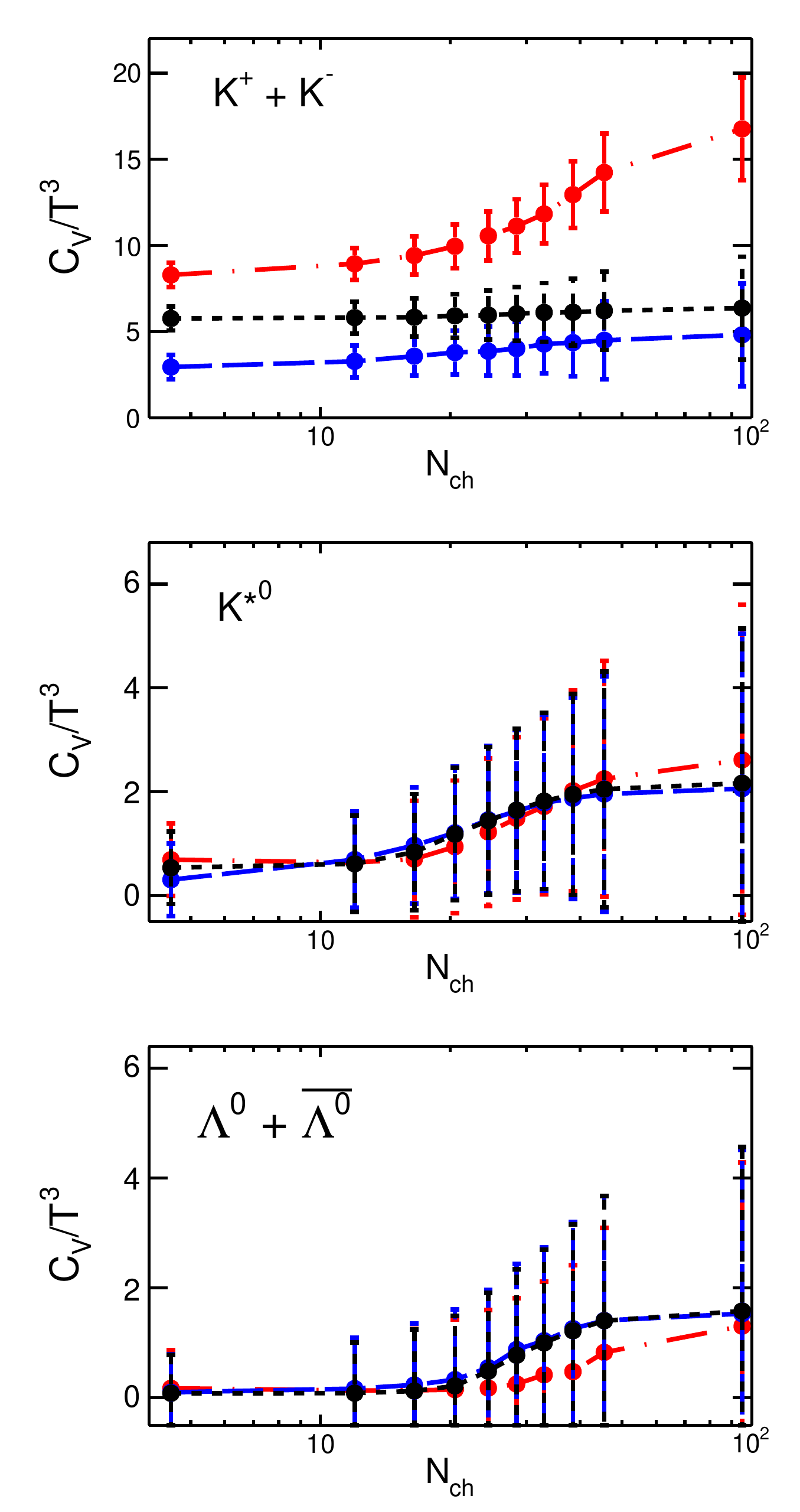}
\caption{(Color Online) Heat Capacity (left), heat capacity scaled by average number of particles (middle) and heat capacity scaled by $T^3$ of the system (right) obtained using Eq.~\ref{eq7} as a function of average charged particle multiplicity for different event shapes for identified strange particles.}
 \label{fig3}  
\end{figure*}

\subsection{Event shape and multiplicity dependence of heat capacity ($C_{V}$)}
Heat capacity of a system is the amount of heat energy required to raise the temperature of the system by
one unit. It can be measured experimentally by measuring the energy supplied to the system and resultant
change in temperature. It gives the measure of how change in temperature changes the entropy of a system 
$(\Delta S =\int  (C_V/T)~dT)$. The change in entropy is a good observable for studying the phase transition. 
In the context of heavy-ion collisions, it can be connected to the rapidity ($y$) distribution ($dN/dy \approx dS/dy$). 
The heat capacity acts as a bridging observable for experimental measurement and theoretical models, where 
change in entropy can be estimated. The heat capacity represents the ease of randomization for a particular phase 
of the matter in opposition to strength of correlation. The scaled value, $C_ V/\langle N \rangle$ remains constant 
with temperature for an ideal gas, since the increase of temperature has no effect on change in interaction strength 
and its range. The heat capacity will change with  some macroscopic conditions if that condition causes changes in 
the strength of correlation and then ease of randomization.  So heat capacity is a good observable to understand 
how correlation and randomization compete over one another. Thus the study of variation of heat capacity with 
multiplicity in pp collisions gives opportunities to have a better understanding of how the ease of randomization 
and the strength of correlation change with number of constituents in a QCD system.

\begin{figure*}[ht!]
\includegraphics[scale=0.45]{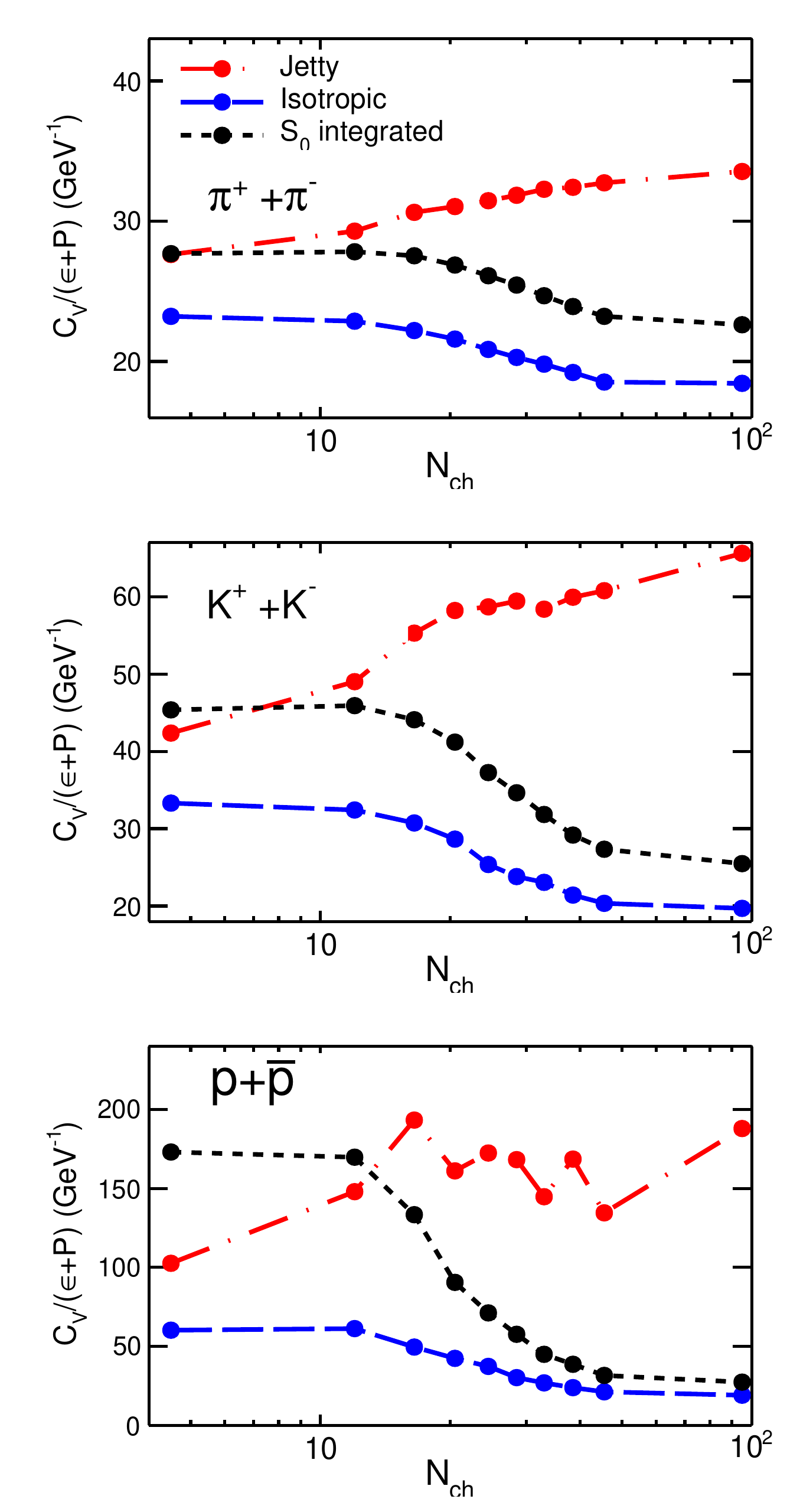}
\includegraphics[scale=0.45]{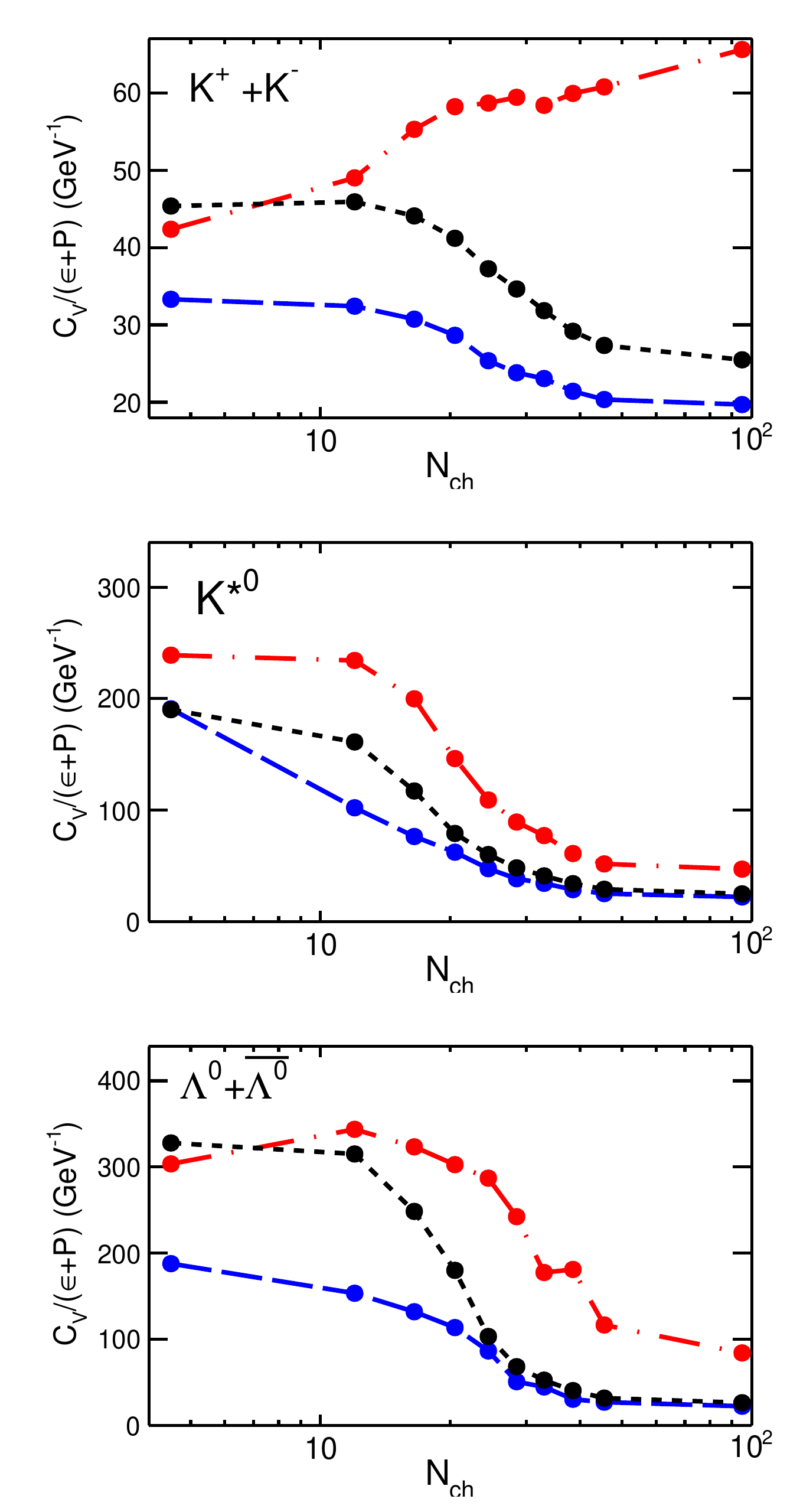}
\caption{(Color Online) Heat capacity scaled by inertial mass of respective light flavor (left) and strange (right) particles 
as a function of event multiplicity and spherocity.}
\label{fig4} 
\end{figure*}
As different event shapes have got different underlying physical mechanisms, it is worth making a comprehensive study of some of the
important thermodynamic observables as a function of event topology through particle spectra in pp collisions using PYTHIA8.
Left panel of Fig.~\ref{fig2} shows the $C_{V}$ of pions, kaons and protons obtained from Eq.~\ref{eq7} using PYTHIA8 simulated data as a function of charged particle multiplicity for different spherocity classes. The lighter mass particles have higher heat capacity, which can be understood from the fact that the production cross-section decreases as a function of particle mass. 
It is also observed that the trend of $C_{V}$ for isotropic and spherocity integrated events are similar and they tend to increase as a function of charged particle multiplicity. At low multiplicity classes, the trend of $C_{V}$ remain almost similar for different spherocity classes. However, the $C_{V}$ for jetty events are always less than the isotropic and spherocity integrated events for high multiplicity classes. This behavior goes inline with our general expectation for the following reasons. It is expected that for the isotropic events, the number of produced particles would be higher compared to that of the jetty events. Thus one would need higher energy to increase one unit of temperature in isotropic events compared to that of jetty ones. As heat capacity is a measure of the amount of energy/heat required to increase one unit of temperature of the system, the isotropic events should have higher heat capacity compared to the jetty ones. As the spherocity integrated events are the average of both isotropic and jetty events, the heat capacity remains in between of the isotropic and jetty events. As, the number of particles seems to play important role in heat capacity, it is worthwhile to look at heat capacity scaled with average number of the corresponding particles under study, which is shown in the middle panel of Fig.~\ref{fig2}. In this case, we observe completely opposite behaviour of heat capacity for isotropic and jetty events. This confirms that the number of particles in a system plays a crucial role for the heat capacity. This behavior is supported by the results of final state multiplicity driving the particle production at the LHC energies~\cite{Tripathy:2019flj,Tripathy:2018ehz}. However, protons behave differently at low multiplicity classes. The right panel of Fig.~\ref{fig2} shows heat capacity scaled with $T^3$, which makes the quantity dimensionless. The $C_{V}/T^3$ increases as a function of charged particle multiplicity for both isotropic and jetty events but the values for pions and kaons are lower for isotropic compared to jetty events. This suggests that the freeze-out temperature and average number of particles play significant role in the values of heat capacity. However, for protons the $C_{V}/T^3$ values seem consistent with each other for different spherocity classes within uncertainties.

\begin{figure*}[ht!]
\includegraphics[scale=0.47]{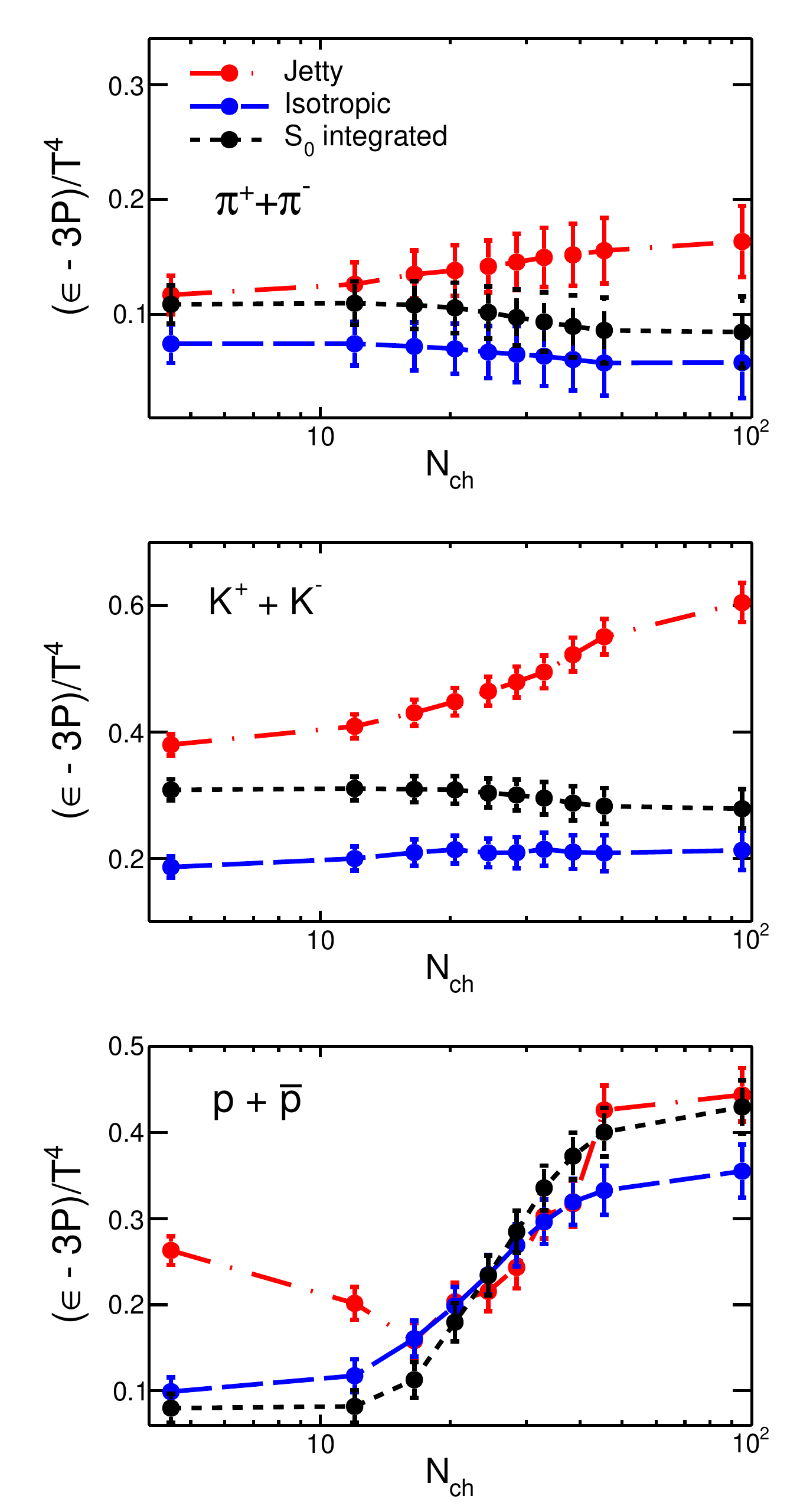}
\includegraphics[scale=0.47]{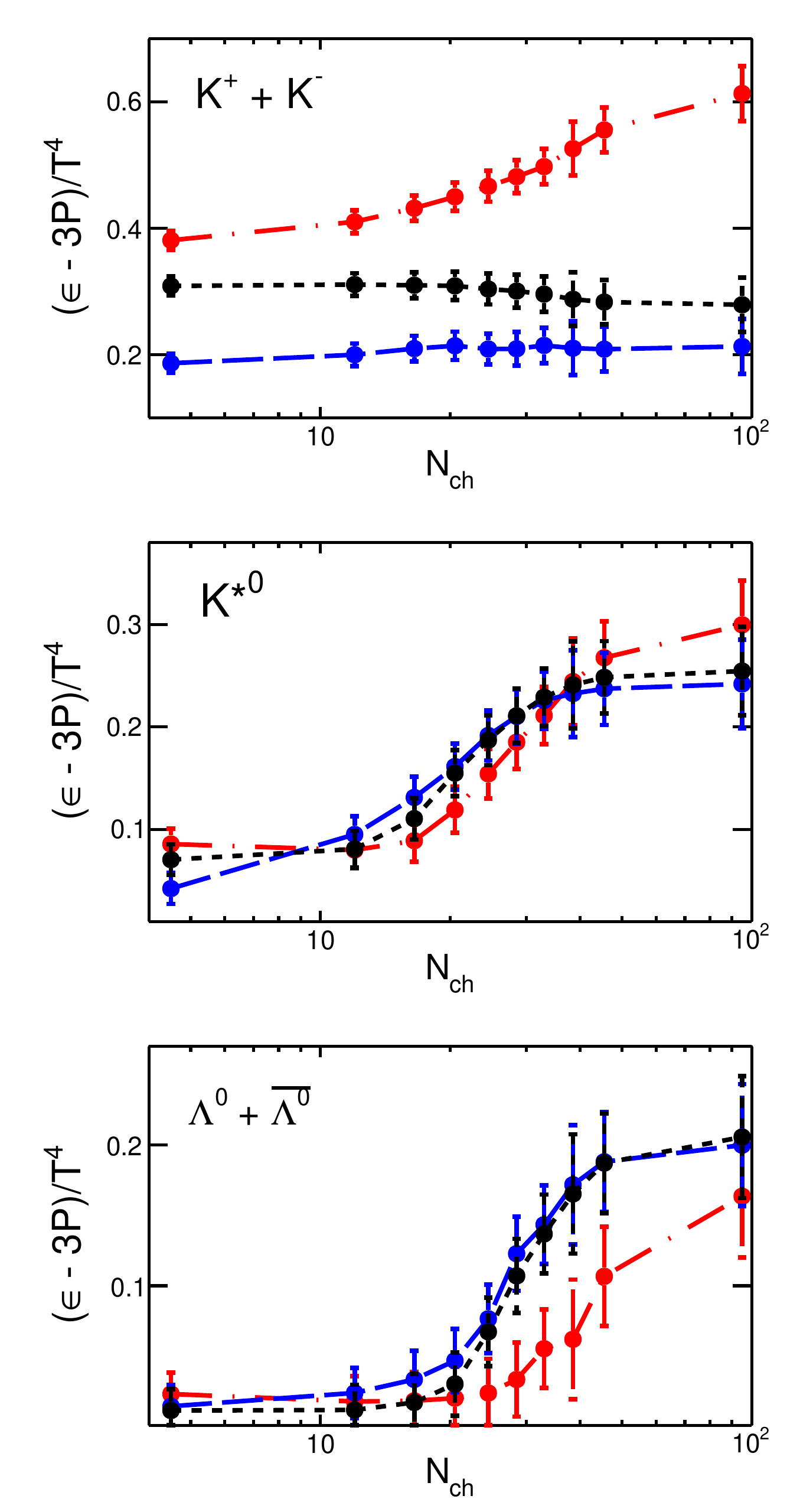}
\caption{(Color Online) CSBM (trace anomaly) for identified light (left) and strange (right) hadrons as a function of event multiplicity and spherocity.}
\label{fig5}
\end{figure*}

Let us now focus on the results from strange particles as it has major significance due to the recent finding of strangeness enhancement in small collision systems like pp and p-Pb collisions~\cite{ALICE:2017jyt}. Figure~\ref{fig3} shows $C_{V}$ (left), $C_{V}/<N>$ (middle) and $C_{V}/T^3$ (right) of strange particles such as kaons, $\mathrm{K}^{*0}$ and $\Lambda^{0}$ obtained from Eq.~\ref{eq7} as a function of charged particle multiplicity for different spherocity classes. We have chosen these particles for our study due to the fact that kaons are the lightest strange mesons, $\Lambda^0$ particles are lightest strange baryons and $\mathrm{K}^{*0}$ are the strange resonances which go through significant re-scattering processes in hadronic phase of the heavy-ion collisions~\cite{Tripathy:2018ehz,Sahu:2019tch}. The behaviours of heat capacity for strange particles are similar to that observed for pions and protons. However, when they are scaled with average number of particles, they show very different behavior compared to pions. The behavior of $\mathrm{K}^{*0}$ and $\Lambda^0$ are similar to that of protons. It is well known that, for a system with finite flow would follow a mass dependent particle production and the thermodynamic observables would be mass dependent. Keeping this in mind, one can expect the similar behaviours of scaled heat capacity with number of particles or temperature for particles with similar masses. As protons, $\mathrm{K}^{*0}$ and $\Lambda^0$ have similar masses the behavior of scaled heat capacity seems to be similar for high multiplicity pp collisions. As seen in the right panel of Fig.~\ref{fig3}, the values of $C_{V}/T^3$ for $\mathrm{K}^{*0}$ and $\Lambda^0$ in different spherocity classes seem to be consistent with each other within uncertainties.

The $(\epsilon+p)$, enthalpy density acts as inertia for change in velocity for a fluid cell in thermal equilibrium. For completeness, we have also studied $C_V$ scaled by enthalpy $(\epsilon+p)$, which acts as a proxy to heat capacity i.e, $C_V$ per unit mass. The specific heat for different spherocity classes as a function of multiplicity for identified stable (left panel) and strange particles (right panel) are shown in Fig~\ref{fig4}. The specific heat seems to have opposite trend to that of heat capacity for all the particles. Also, there is no significant differences of specific heat for different particles as a function of multiplicity and spherocity. It is to be noted here that the behaviour of heavier hadrons like proton, K$^{*0}$ and $\Lambda^{0}$ are quite similar except for jetty events in case of proton.  These hadrons seem to have $S_{0}$ and isotropic events overlap beyond $\rm N_{ch} \simeq$ (20-30). This is expected as heavier hadrons have relatively smaller abundances.

\subsection{Event shape and multiplicity dependence of CSBM, speed of sound}
 \label{mult_CSB_C_s_2}
 Speed of sound in a system reveals about the strength of interactions of the constituents of a medium. A comparison with the standard
 massless ideal gas value would give a hint about the system dynamics. The effective mass of the constituents can change in the presence 
 of interaction, which changes the speed of sound in a medium. The measure of deviation from masslessness of the constituents is captured by CSBM (how particle mass and temperature contributes to CSBM for non-interacting (ideal gas) system is 
 discussed in Ref. \cite{Deb:2019yjo,Sarwar:2015irq}. For massless particles, $c_s^2 = 1/3$. However for massive particles, it is less than 
 this value. This is because, massive particles do not contribute to the pressure as much as they contribute to the energy of a system. It is expected that the variation of these quantities with event multiplicity will capture the change in effective interaction among the constituents with increase in number of constituents. It becomes important to study these quantities as a function of event topology, as topology is a consequence of the underlying particle production mechanism.

Therefore, we have also studied the conformal symmetry breaking measure (CSBM) and squared speed of sound ($c_s^2$) as a function of multiplicity and spherocity for identified particles in pp collisions, which can be obtained using equations Eq.~\ref{eq8},~\ref{eq9}. 
Figure~\ref{fig5} shows CSBM ($\frac{\epsilon - 3P}{T^4}$) of identified stable (left panel) and strange (right panel) particles using T and $q$ obtained from PYTHIA8 as a function of charged particle multiplicity for different spherocity classes. It is observed that the CSBM increases with increase of mass. For spherocity integrated events, the trace anomaly remains almost flat as a function of multiplicity for pions and kaons while it increases for heavier mass particles like protons, $\mathrm{K}^{*0}$ and $\Lambda^0$ particles. For pions and kaons the CSBM is higher for jetty events compared to isotropic events throughout all the charged particle multiplicity classes. However, for other heavier particles CSBM seems to be similar within uncertainties for different spherocity classes in high multiplicity pp collisions.

\begin{figure*}[ht!]
\includegraphics[scale=0.43]{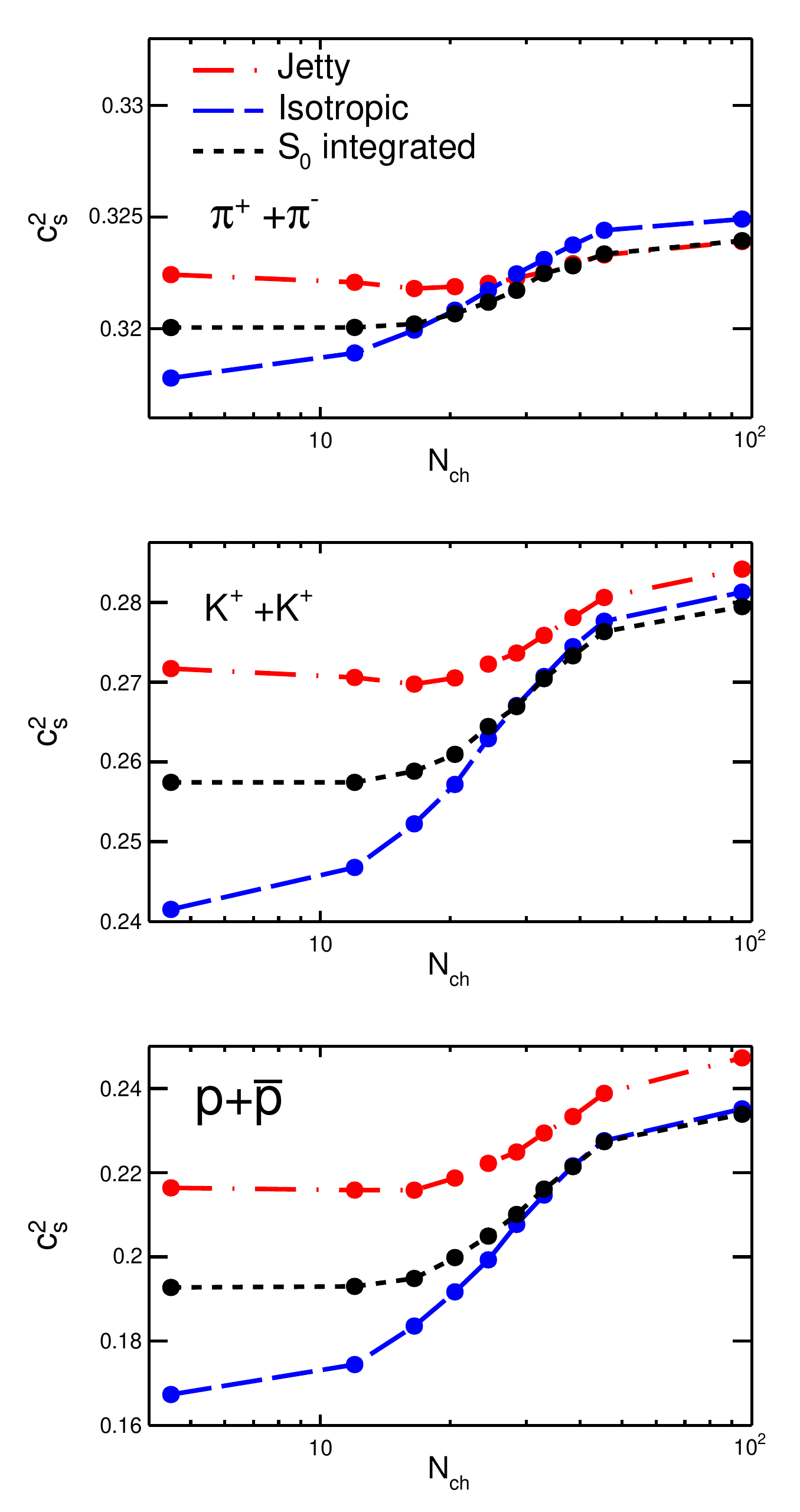}
\includegraphics[scale=0.43]{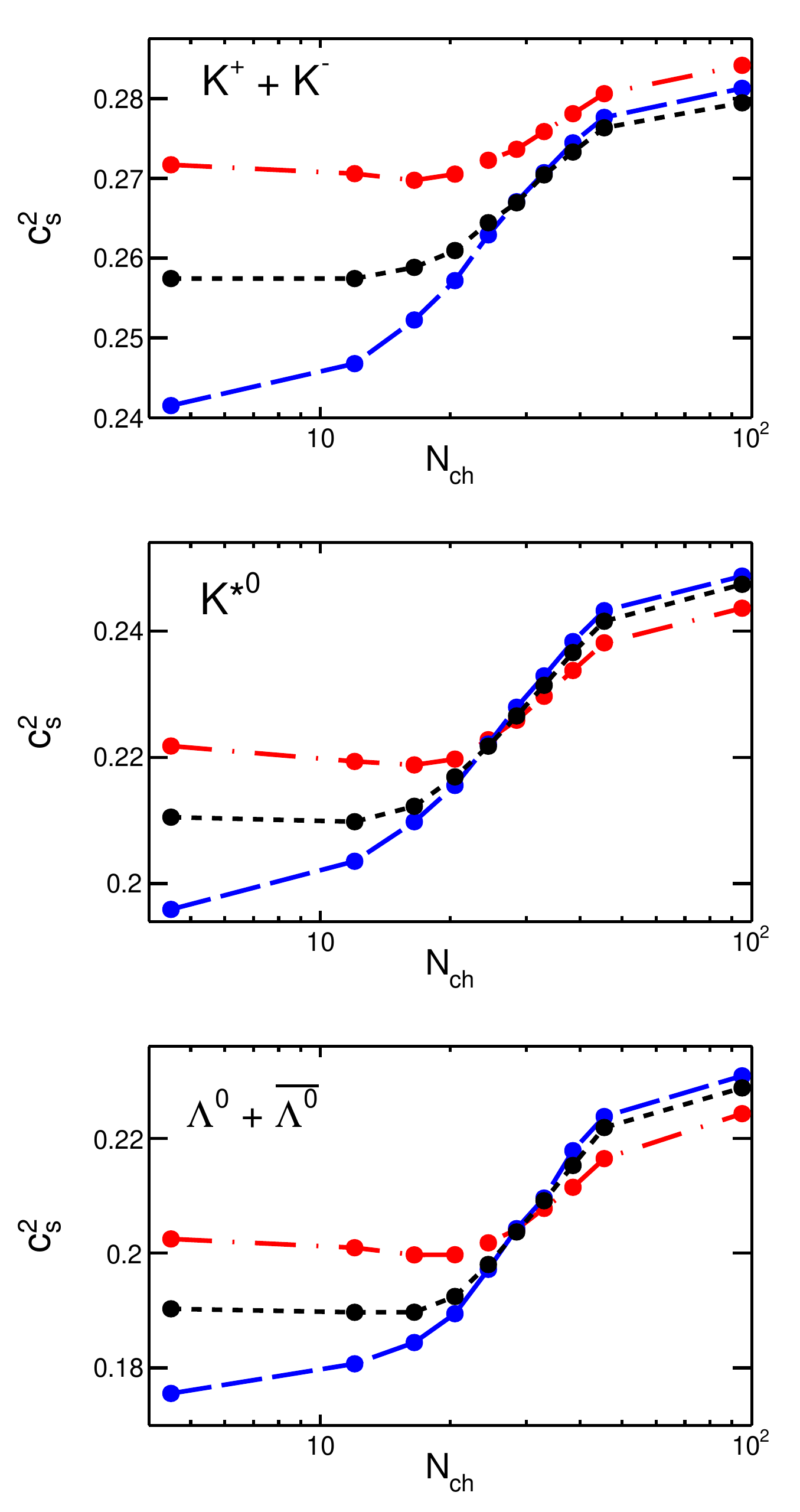}
\caption{(Color Online) Squared speed of sound for identified light (left) and strange (right) hadrons as a function of event
multiplicity and spherocity.}
\label{fig6} 
\end{figure*}

Figure~\ref{fig6} shows the squared speed of sound, $c_{s}^{2}$ of identified stable (left panel) and strange (right panel) particles using T and $q$ obtained from PYTHIA8 as a function of charged particle multiplicity and spherocity. The $c_{s}^{2}$ seems to be mass dependent and decreases with increase in particle mass. Contrary to the other observables, the trend of $c_{s}^{2}$ for different spherocity classes as a function of multiplicity for all the particles are similar and seems to approach the Stefan-Boltzmann limit of $1/3$, asymptotically. This behavior is consistent with our 
earlier work~\cite{Deb:2019yjo}.

For all the above discussed thermodynamic observables, a common feature 
appears, which is the threshold in final state event multiplicity. 
The system behavior changes for the value of final state event 
multiplicity more than 
$\rm N_{ch} \simeq (10-20)$. This is a confirmatory observation as a threshold final state event multiplicity in high-multiplicity pp collisions. This goes inline with many such earlier observations of a threshold in final state event multiplicity after which MPI shows substantial activity and explains charmonia production \cite{Thakur:2017kpv}, thermodynamic limit of all the statistical ensembles showing similar freeze-out properties \cite{Sharma:2018jqf} and the saturation of non-extensive thermodynamical parameters \cite{Rath:2019cpe}.  
Further it should be noted here, that as an emerging area of final state multiplicity driving the multiparticle production
processes in hadronic and nuclear collisions at the LHC energies, although systematic study taking the final state multiplicity becomes evident, so far the thermodynamics of the system is concerned, the physical interpretation of observables for smaller number density should be taken with caution. We believe, the present work along with many others in the direction of event topology dependent studies at the LHC energies are a way forward in 
understanding the heavy-ion-like features in high-multiplicity pp collisions and a possible formation of QGP-droplets \cite{Sahoo:2020cle,Sahoo:2019ifs}. These aspects should have an experimental exploration, once the corresponding data become available. This study, thus paves a way to understand the high-multiplicity pp collisions at the LHC energies.

 \section{Summary}
  \label{sum}
  In this work, we have made an attempt to study event topology and event multiplicity dependence of some of the
  important thermodynamics variables in pp collisions at $\sqrt{s}$ = 13 TeV in view of the heavy-ion-like features observed in high-multiplicity
  events. In the absence of transverse spherocity analysis in experimental data, we have used pQCD inspired PYTHIA8 event generator to simulate pp collisions at $\sqrt{s}$ = 13 TeV. As the production dynamics of hard QCD and soft processes contribute differently to the event structures, we have used transverse spherocity as an event topology analysis tool to separate jetty and isotropic events, and then study some of the important thermodynamic observables. It is quite evident from the above observations that the results for spherocity integrated events fall in between of isotropic and jetty events. This suggests that the spherocity plays a significant role of separating events based on their geometrical shapes. This also indicates that studying  all the events without looking at the geometrical shape of the events might not contain the entire information about the possible flow-like medium and/or jets. Also, one can notice from all the results that there is a threshold number of charged particles after which the behavior of the observables changes significantly in isotropic, jetty and spherocity integrated events. This threshold is found to be $\rm N_{ch} \simeq$ (10-20), and becomes an important and confirmatory finding over earlier such observations~\cite{Thakur:2017kpv,Campanini:2011bj}. In general, in a many particle system the lighter particles predominantly contribute to its thermodynamic properties. In the present context, pions and kaons are lighter as compared to other hadrons considered. Hence these particles govern the thermodynamic behaviour of the system with their higher abundances. The variations of thermodynamical quantities considered in present work with $\rm N_{ch}$ for lighter hadrons like pions and kaons show a plateau which starts at a low value of $\rm N_{ch}$. We find that a similar plateau-like behaviour is also achieved for heavier hadrons, like proton, K$^{*0}$ and $\Lambda^{0}$ for $\rm \langle N_{ch} \rangle$ $>$ 40, indicating a scenario where a thermal bath has been formed  with all these hadrons in equilibrium. As heavier hadron are relatively less abundant, it is expected to form a thermal bath for higher $\rm N_{ch}$ than the lighter hadrons like pions and kaons. We believe such a study based on pQCD inspired PYTHIA8 event generator using event topology and multiplicity becomes important in exploring the production dynamics of high-multiplicity pp collisions. 
  
 It should be noted here that in PYTHIA8, a partonic medium is not explicitly invoked. Rather, MPI with color reconnection has been 
 successful in describing the collectivity observed in pp collisions at the LHC energies. The present observation of a threshold in the particle
 multiplicity indicating a dynamical behaviour in particle production and the thermodynamics of the produced system is a consequence of MPI with color reconnection.
  
 Further to support similar observations of event multiplicity threshold, of charged particle density in pseudorapidity greater than (10-20), in a parallel work
 by one of us \cite{me},  in a color string percolation scenario using the experimental particle spectra in hadronic and nuclear collisions at 
 various LHC energies the following observations are made. For multiplicity density greater than 20,
it is observed that the required critical initial energy density ($\epsilon_c \sim 1 ~ {\rm GeV/fm^3}$) and the critical deconfinement/hadronization temperature ($T_c \sim $ 167.7 $\pm$ 2.8 MeV) are achieved. In addition, the shear viscosity to entropy ratio ($\eta/s$) studied as a 
function of event multiplicity shows comparable values of $\eta/s$ in high-multiplicity pp events \cite{me,Hirsch:2018pqm} and heavy-ion collisions, revealing the formation of a strongly coupled QGP in collisions for which the multiplicity density is greater than 20. These observations have profound implications in the search for QGP-droplets in high-multiplicity pp collisions and characterization of the
system produced in heavy ion collisions at relativistic energies.

\section{Acknowledgement} 
This work is supported by DAE-BRNS Project No. 58/14/29/2019-BRNS of R.S. and the DAE Principal Collaborator, J.A. The authors would like to thank Mr. Pritindra Bhowmick of IISER Bhopal for initial exploration of the present work during his short summer internship at the Indian Institute of Technology Indore, India.


 \end{document}